\documentclass{iau_FM}
\usepackage{graphicx}
\usepackage{natbib}
\usepackage{aas_macros}
\usepackage{amsmath}
\usepackage{hyperref}

\usepackage{setspace}
\title[Revisiting turbulent properties of solar convection] %% give here short title %%
{Revisiting turbulent properties\\of solar convection with\\3D radiative hydrodynamic modeling}

\author[Irina N. Kitiashvili, Alan A. Wray]   %% give here short author list %%
{Irina N. Kitiashvili$^1$, Alan A. Wray$^2$}
\affiliation{NASA Advanced Supercomputing Division, NASA Ames Research Center, \\
N258, Moffett Field, CA 94035, United States  \\ 
email$^1$: {\tt irina.n.kitiashvili@nasa.gov}\\
email$^2$: {\tt alan.a.wray@nasa.gov}}

\pubyear{2024}
\jname{Focus Meeting 8: Advances and Challenges\\in Understanding the Solar and Stellar Dynamos} 
\editors{A.C. Editor, B.D. Editor \& C.E. Editor, eds.}
\begin{document}

\maketitle

\begin{abstract}
We discuss the turbulent structure and dynamics of the upper solar convection zone using a 3D radiative hydrodynamic simulation model at 45 degrees latitude. The model reveals the self-formation of meridional flows, the leptocline, and the radial differential rotation. Unlike previous studies, the model shows a complex variation of the characteristic scales of turbulent flows with depth. 
In particular, an increase in the characteristic convective scale is trackable within an individual snapshot up to a depth of 7~Mm, near the bottom of the hydrogen ionization zone, where turbulent flows become weaker and more homogeneous. However, the turbulent spectra show an increase in scale with depth and a qualitative change in convective patterns below 7~Mm (near the bottom of the leptocline), suggesting changes in the diffusivity properties and energy exchange among different scales.
%% add here a maximum of 10 keywords to be taken form the file <Keywords.txt>
\end{abstract}

\firstsection % if your document starts with a section,
              
\section{Introduction}
It is well known that solar and stellar variability are tightly connected to the properties and dynamics of turbulent flows in the convection zone. Most of the current understanding of the structure and dynamics of the convection zone of the Sun comes from photosphere observations and 3D modeling on local and global scales. Because there are no direct observations of the solar interior, information about subsurface flows is mainly retrieved from observations of photospheric disturbances, such as tracking of various features \citep[e.g.,][]{Svanda2013,Hathaway2021} and analysis of oscillations in the photosphere using helioseismology methods \citep[e.g.,][]{ChristensenDalsgaard2002,Kosovichev2011}.
In particular, we have a general understanding of how subsurface flows change over solar activity cycles and how it may be linked to observed magnetic flux evolution on the solar surface, both on global \citep[e.g.,][]{Howe2017,Getling2021,Komm2021} and local scales \citep[e.g.,][]{Jain2016,Stefan2023,Braun2024,RabelloSoares2024}. In addition, available computational capabilities allow for realistic modeling of solar dynamics on global and local scales. 

Recent progress in 3D global modeling has shown the ability to qualitatively reproduce a solar-type differential rotation, with a faster rotation at the equator and slower at high latitudes \citep[e.g.,][]{Guerrero2013,Matilsky2019,Hotta2022}. 
However, many details of the global dynamics still deviate from observations. The problem of modeling solar subsurface dynamics depends on the accuracy of modeling turbulence in the convection zone. Significant progress has been made in understanding the essential ingredients needed for realistic modeling of solar convection on global scales, such as resolution requirements and turbulent transport coefficients \citep[e.g.,][]{Penev2007,Karak2018,Warnecke2020,Hotta2022,Gupta2023}, and the phenomena of convective overshoot, magnetic fields, and rotation \citep[e.g.,][]{Guerrero2013,Fan2014,Beaudoin2018,Kaepylae2019}, as well as others. 

In contrast to the solar interiors, the solar photosphere and low atmosphere have been observed in great detail with ground and space instruments at different wavelengths.  These observations aided the development of 3D realistic-type modeling based on ab initio or first physical principles. This approach has demonstrated the capability of such modeling to reproduce and understand processes behind many observed phenomena \citep[e.g.,][]{Stein2001,Nordlund2009,Rempel2009,Kitiashvili2010,Kitiashvili2013,Bjoergen2019,Cheung2019,Chen2023}. 

Ongoing developments of 3D numerical modeling, which take into account the effects of subgrid-scale turbulence, have already demonstrated the crucial role of the shallow, $\sim 8$~Mm deep leptocline layer (`the leptocline') associated with hydrogen and helium ionization zones \citep{Kitiashvili2023}.  In the present paper, we revisit the problem of the change in convection structure with depth and examine the convective properties in the upper 20~Mm of the solar convection zone in the presence of rotation with the Coriolis force corresponding to 45~degrees latitude. 

\section{Model setup}

We analyze a computational model similar to the 3D radiative hydrodynamic simulations by \cite{Kitiashvili2023} with the `StellarBox' code \citep{Wray2015,Wray2018}. In this simulation, the modeling of solar convection is performed for the realistic rotation rate that corresponds to a 45-degree latitude retrieved from a global helioseismology data product of the SDO/HMI pipeline (available from the Joint Science Operations Center, JSOC \footnote{JSOC: \url{http://jsoc.stanford.edu}}). The imposed uniform rotation rate of the computational domain corresponds to the mean solar rotation at a 45-degree latitude from --25~Mm to --2~Mm below the photosphere, averaged over 3 months of observations during the deep solar minimum in 2008. To investigate the subsurface structure of the modeled solar convection, we utilize 3D hydrodynamic data from a 1254-hour-long computational run to ensure that a statistically stationary state of turbulent convection is well approximated. In this paper, we analyze the last 24 hours of these computations.

\begin{figure}%[h]
% \vspace*{-2.0 cm}
\begin{center}
 \includegraphics[width=4.35in]{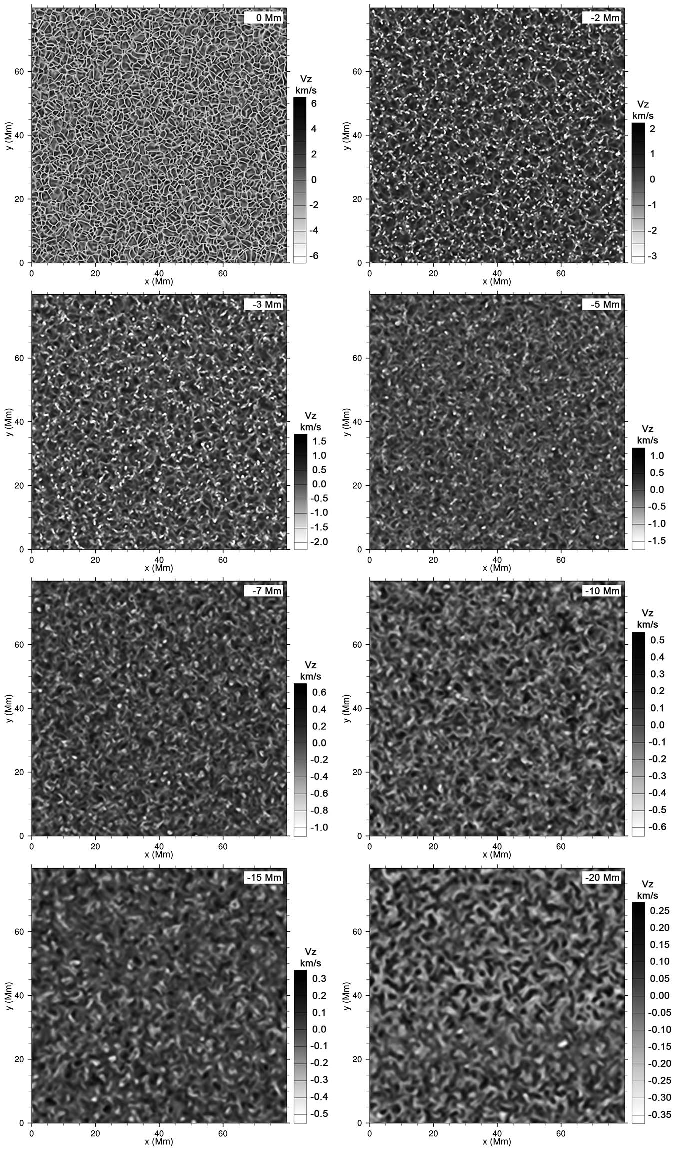} 
% \vspace*{-1.0 cm}
 \caption{
 Snapshots of the radial (vertical) velocity from the 3D radiative hydrodynamic model for 8 layers in the solar convection zone at 0 Mm (photosphere), --2 Mm, --3 Mm, --5 Mm, --7 Mm, --10 Mm, --15 Mm, and --20 Mm below the surface, showing qualitative changes of turbulent convection. Brighter colors correspond to the downflows.}
   \label{Vz_xy}
\end{center}
\end{figure}

\section{Subsurface structure of upper convection zone}

The convective structure at the photosphere is primarily represented by granulation with a characteristic size of 1 -- 2~Mm. In the subsurface layers, the structure of the convection and its characteristic scales change with depth.
Figure~\ref{Vz_xy} shows the vertical velocity distribution from the photosphere, z = 0~Mm, to 20~Mm below. Black colors in this figure correspond to upflows, and white colors correspond to downflows. In the photosphere, granulation is easily distinguishable as dark gray islands. Below the photosphere, at --2~Mm (granulation layer), there are no clear intergranular lanes; the scale of the convection is primarily dependent on the characteristic distance between individual downdrafts. As we go to still deeper layers, the characteristic scale increases further as fewer downdrafts can penetrate these deeper layers, which agrees with previous studies by \cite{Nordlund2009}. 

At a depth of 7~Mm, the clustering of the downdrafts and larger structures is not obvious but still can be identified (Fig.~\ref{Vz_xy}). The flow structure at this depth represents a mixture of upflows and downflows, which prevents us from identifying large patterns from a single snapshot and requires time-averaging for such identification. Such identification contradicts the previous findings of \cite{Nordlund2009}, in which the subsurface flows between the downdrafts were significantly less turbulent. This difference can be attributed to use of the Smagorinsky subgrid-scale turbulence model, modified for compressible flows \citep{Smagorinsky1963,Moin1991,Germano1991}.
Thus, in these deeper layers, the turbulent motions in the individual snapshots show an overall decrease in the amplitude of the flows and changes in structure, with the result that a purely visual identification of scale sizes becomes difficult.

\begin{figure}[b]
	% \vspace*{-2.0 cm}
	\begin{center}
		\includegraphics[width=5.2in]{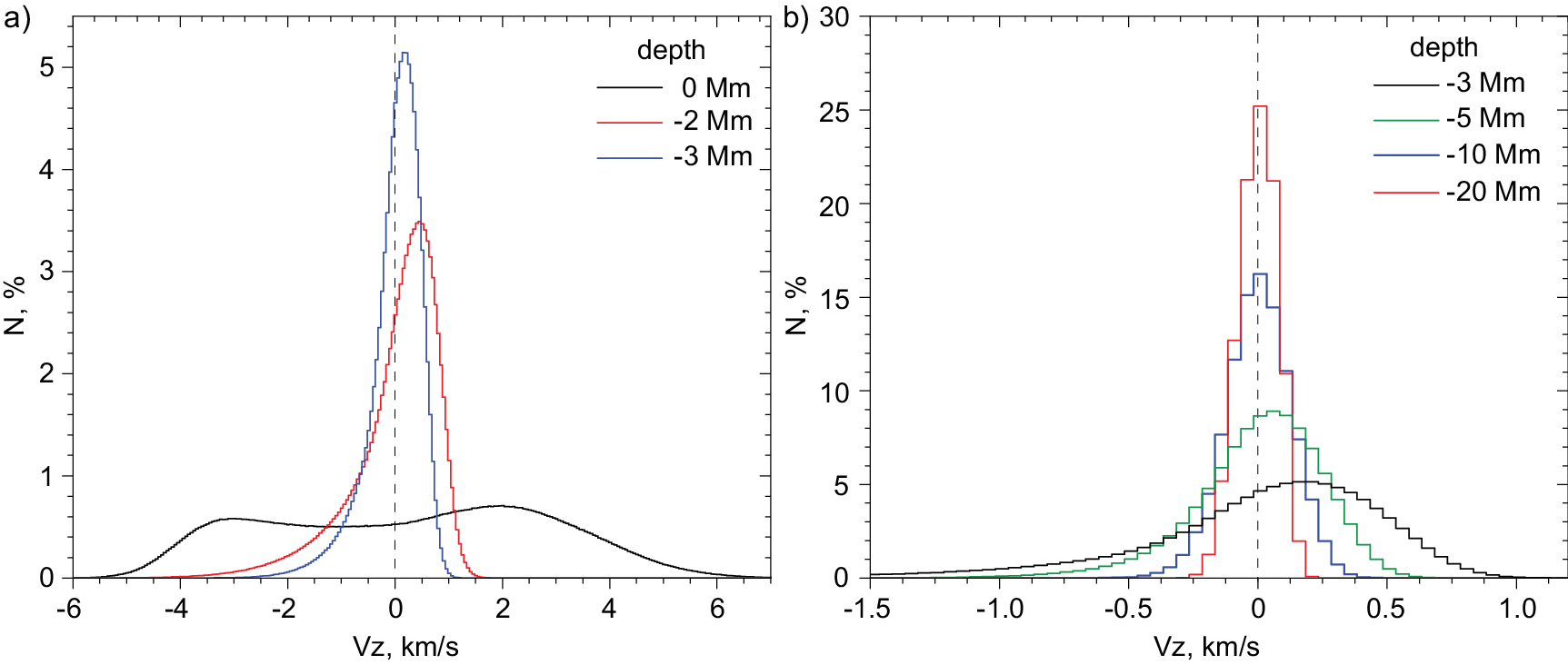} 
		% \vspace*{-1.0 cm}
		\caption{The histograms show a distribution of the vertical velocity, Vz, in the six layers shown in Figure~\ref{Vz_xy} of the solar convection zone from the photosphere to 20~Mm below. The histograms are computed from 24 snapshots separated by 1~hour. The bin size is 50~m/s.}
		\label{Vz_hist}
	\end{center}
\end{figure}

We show the statistical distribution of the radial component of the convective velocity in Figure~\ref{Vz_hist}. As expected, it reveals a dramatic difference between photospheric and subsurface flows (Fig.~\ref{Vz_hist}a). At the photosphere (black curve, panel a), the two bumps correspond to characteristic upflows of granulation ($\sim 2$~km/s) and downflows in intergranular lanes ($\sim -5$~km/s). At the bottom of the granulation layer (2 -- 3~Mm deep; red and blue curves), the histograms show dominant downward motions driven by radiative cooling at the photosphere. Since the granulation layer of the quiet Sun extends down to 2 -- 3~Mm and contains significant strong downdrafts, the contrast between upflows and downflows is highest there (Fig.~\ref{Vz_hist}a), allowing identification of the convective scales visually from a snapshot. Because downflows cannot penetrate deep into the solar interior \citep[e.g.,][]{Nordlund2009,Kitiashvili2019,Kitiashvili2023}, the statistical distribution between upflows and downflows becomes closer to a Gaussian one (Fig.~\ref{Vz_hist}b) and prevents us from visually tracking the convective scale increase into deeper layers. 

\begin{figure}[t]
% \vspace*{-2.0 cm}
\begin{center}
 \includegraphics[width=3.in]{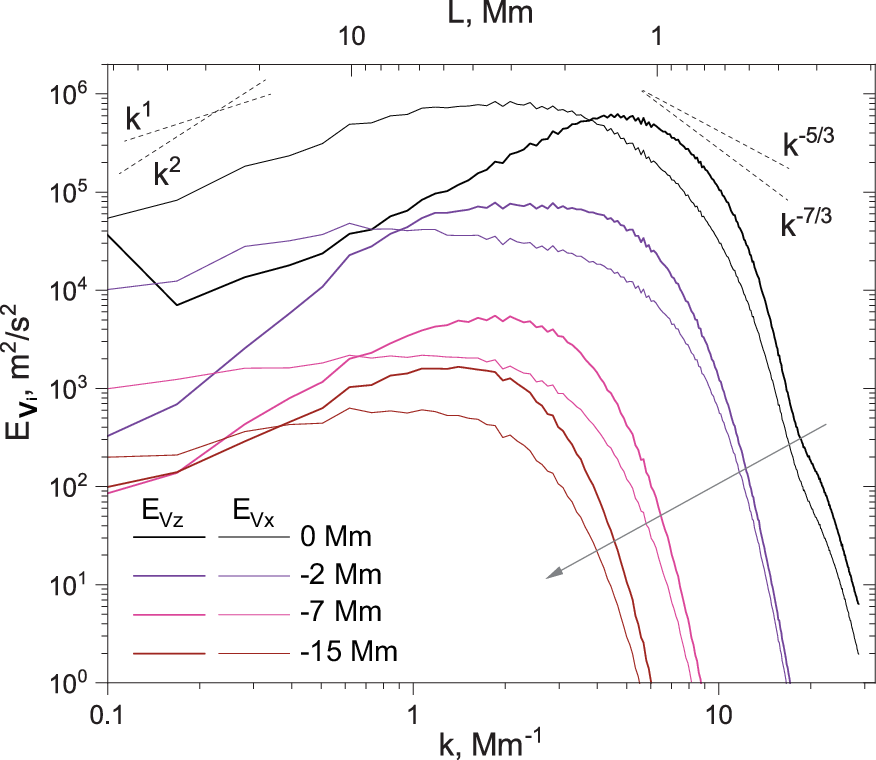} 
% \vspace*{-1.0 cm}
 \caption{Turbulent spectra of the radial (Vz, thick curves) and azimuthal (Vx, thin curves) components of velocity at different depths. In the plot, `k' is the horizontal wavenumber, and L=2 $\pi$/k. The spectra were obtained from 24 snapshots separated by 1~hour. The gray arrow indicates the direction of change in the spectra at depths from the surface to the interior.}
   \label{spectra}
\end{center}
\end{figure}

To identify the characteristic turbulent scales, it is natural to analyze the simulation data in terms of the spectral energy. To track the scale change with depth, we calculate turbulent spectra for vertical (Vz) and azimuthal (Vx) velocities (Fig.~\ref{spectra}).  
For the vertical velocity, the granulation scale can be easily identified at the photosphere as a bump at L$\sim 1.5$~Mm (k $\sim 4 $~Mm$^{-1}$). This bump in the turbulent spectra becomes wider and shifts to larger scales as depth increases, from $\sim 1.5$~Mm at the photosphere (black thick curve) to 3 -- 9~Mm at 15~Mm below the surface (thick brown curve). 

At the photosphere, narrow ranges of the vertical velocity spectra show power-law indices (the sloping lines in Fig.~\ref{spectra}) that correspond to the classical Kolmogorov slope of $k^{-5/3}$. This slope is also present, at larger scales, in the deeper layers. 

For the horizontal flows, the power index corresponds to a slope closer to $k^{-7/3}$ at the photosphere and changes with depth, becoming closer to $k^{-5/3}$ in deeper layers. 
At the photosphere, the spectra for the azimuthal component of the flows show that the energy is mainly concentrated at scales of 2.5 -- 8~Mm. There are no distinct characteristic scales below the photosphere. The turbulent spectra for the meridional flow component (Vy) have the same properties as the spectra for the azimuthal component (Vx). The azimuthal and meridional components of the flows deviate on scales larger than 10~Mm.

The identified slope of $k^{-5/3}$ in a small wavenumber range of the turbulent spectra indicates that the turbulent plasma in the solar interior is approximately homogeneous at those length scales. However, it is important to note that the \cite{Kolmogorov1941} theory of turbulence was developed for idealized homogeneous turbulence, where many effects, such as radiative energy transport and strong stratification, are not present. Therefore, the solar turbulence spectrum cannot, on this evidence alone, be proven to have a Kolmogorov-type inertial range \citep{Kitiashvili2013a}. Of course, additional inhomogeneity will occur in the presence of magnetic fields. 

\section{Impact of the convection zone structure on turbulent properties}

\begin{figure}[t]
% \vspace*{-2.0 cm}
\begin{center}
 \includegraphics[width=5.2in]{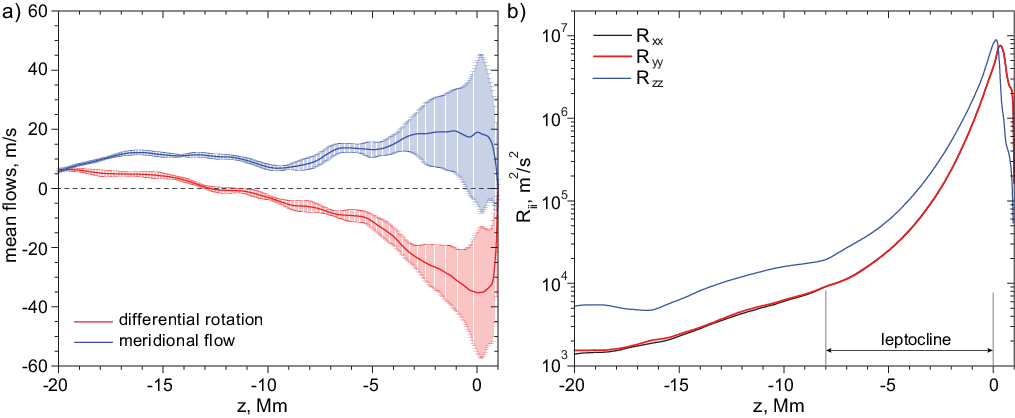} 
% \vspace*{-1.0 cm}
 \caption{Panel a: Radial profiles of the rotational velocity (red curve) and meridional flow velocity (blue). The vertical bars show the standard deviation from the mean. Panel b: The mean radial profiles of diagonal Reynolds stresses. The radial profiles are obtained from 24 hours of the simulation data.}
   \label{flows}
\end{center}
\end{figure}

The significant changes in the turbulent spectra with depth in the upper convection zone raise questions about how these changes affect the mean flows and thermodynamical structure in these layers. Because of the constant rotation rate of the computational domain that corresponds to the mean rotation rate in the upper 25~Mm of the solar convection zone at 45-degree latitude, the impact of the rotation is weaker in comparison to the previous study by \cite{Kitiashvili2023}, where the higher Carrington rotation rate was applied. Similar to the earlier studies, the differential rotation in this case reveals slower rotation of the photospheric flows by about --35~m/s. The rotational velocity of subphotospheric flows increases with depth (red curve, Fig.~\ref{flows}a). The meridional flows (blue curve) are directed Northward with a speed of about 20~m/s from the photosphere to a depth of about $-3$~Mm. Below the granulation layer, the flows are nearly constant, $\sim 10$~m/s. The computations and the subsequent analysis showed that the large-scale flows vary in time on significantly longer (over 24 hours) time-scales considered in this paper. For example, the meridional flows may significantly decrease during some periods and even change direction to equatorward. The structural changes in the large-scale flows (both the differential rotation and meridional flows) are often associated with the bottom of the leptocline layer discovered in helioseismic observations \citep{Deubner1979,Rozelot2009} and in numerical simulations \citep{Kitiashvili2023}. Because the bottom of the leptocline layer, at a depth of about $-8$~Mm, represents an interface between stronger and weaker convectively unstable layers, the diagonal components of the Reynolds stresses show a small dip (for R$_{\rm xx}$ and R$_{\rm yy}$) or bump (for R$_{\rm zz}$; Fig.~\ref{flows}b) there.

\begin{figure}[b]
	% \vspace*{-2.0 cm}
	\begin{center}
		\includegraphics[width=5.3in]{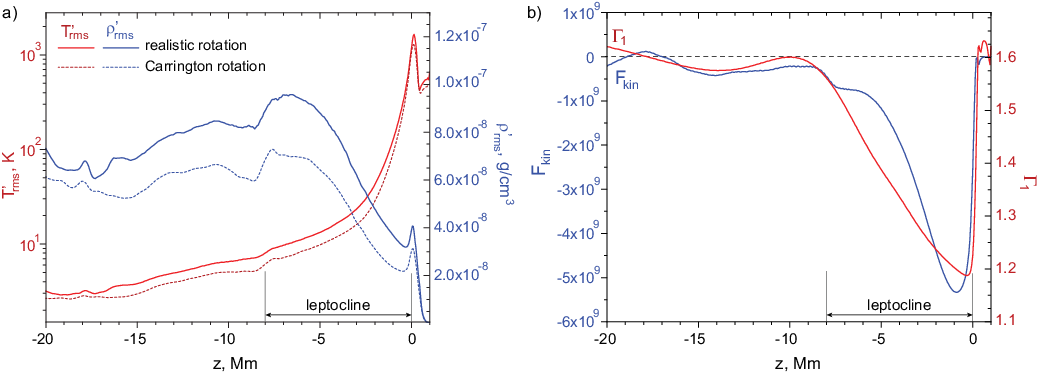} 
		% \vspace*{-1.0 cm}
		\caption{Panel a: Radial profiles of $rms$ fluctuations of temperature (red curve) and density (blue). The solid curves correspond to the realistic rotation, and the dotted curves correspond to the Carrington rotation rate. Panel b: Variation of the kinetic energy flux associated with convection, ${\rm F_{kin}}$ (blue curve) and the adiabatic index, $\Gamma_1$ (red).}
		\label{TRho}
	\end{center}
\end{figure}

The bottom of the leptocline boundary is noticeable in the $rms$ variations of the density and temperature perturbations (Fig.~\ref{TRho}a). Despite the qualitative similarity to the case of faster rotation \citep[Fig.~8a in][]{Kitiashvili2023}, the gradient of temperature perturbations (red curve) is significantly weaker in the simulations with the rotation rate taken from helioseismic inversions. Similarly, in this case, using the realistic rotation rate, the density variations are stronger, and the interface between the bottom boundary of the leptocline and the deeper layers is smoother in comparison to the case of the faster Carrington rotation \citep{Kitiashvili2023}. Variations of the adiabatic index, $\Gamma_1$, and the kinetic energy flux associated with convection  (Fig.~\ref{TRho}b) are co-located with the leptocline layer. This suggests that the hydrogen and helium second ionization zones that cause the $\Gamma_1$ dips at a depth of 0 -- 9~Mm and 11 --16 Mm are significant for structuring flows in the upper convection zone. The kinetic energy flux (blue curve, Fig.~\ref{TRho}b), defined as \citep{Nordlund2001}
\begin{equation*}
	F_k=\left\langle \left( \frac{1}{2}\rho V^2\right) \breve{V}_z\right\rangle_{xy}, \quad \breve{V}_z=V_z-\bar{V}_z, \quad  \bar{V}_z=\left\langle \rho V_z\right\rangle/\left\langle \rho\right\rangle,
\end{equation*}
correlates with the upper layer of the H and He-I ionization zones. The following relatively small increase of the kinetic energy flux around $-7$~Mm (a negative kink of ${\rm F_{kin}}$) corresponds to the bottom boundary of the leptocline. The correlation of the kinetic energy flux with the $\Gamma_1$ in the He-II ionization zone is potentially relevant to an increase in the homogeneity of the turbulent flows in terms of the kinetic energy exchange between different scales. In particular, a comparison of the turbulent spectra obtained for different velocity components shows a strong deviation of the spectral energy distribution on larger scales. In contrast, the turbulent spectra at a depth of $-15$~Mm and the turbulent energy distribution for the radial and horizontal flows (Fig.~\ref{spectra}) at scales above 20~Mm are close to each other. 

\section{Discussion}

The problem of the turbulent structure and dynamics of solar convection is challenging because no direct observations are available. The turbulent properties of the solar interior have been estimated from observations and also through stellar evolution modeling using mixing length theory \citep[e.g.,][]{Bohm-Vitense1958,Cox1968}. These methods suggest a gradual increase in the convective scales with depth over several density scale heights. Thanks to growing computing capabilities, realistic 3D radiative hydrodynamic simulations also demonstrate an increase of the convective scales with depth up to 12~Mm below the photosphere, but further show that the structure of the upper convection zone is more complicated than predicted by those earlier methods. 

In particular, an increase in the convective scales can be identified up to about 7~Mm deep in individual snapshots (Fig.~\ref{Vz_xy}), where the contribution of downflows is more significant in comparison to the upflows (Fig.~\ref{Vz_hist}). That is in agreement with previous numerical studies \citep{Nordlund2009}. However, in the deeper layers, an increase of the convection scales can be only noticed in the turbulent spectra (Fig.~\ref{spectra}) as the amplitude of the radial flows decreases and the relative distribution of the vertical velocities becomes close to Gaussian (Fig.~\ref{Vz_hist}). This result suggests that integration over time is required to capture the convective scale increase.

To investigate a connection between a qualitative change in the turbulent scales and other mean properties of the convection zone, we consider the mean thermodynamic and flow properties in the presented simulations. In particular, the radiative hydrodynamic simulations reveal the formation of differential rotation and meridional flows, and a 10-Mm thick layer, a so-called `leptocline,' previously found in a model for faster rotation rate \citep{Kitiashvili2023}. The leptocline co-locates with the H and He-I ionization zones and is associated with a decrease in the $rms$ of temperature and density perturbations and with variations in the kinetic energy flux (Fig.~\ref{TRho}) that are also responsible for changes in turbulence structure. 
Considering the radial distribution of the diagonal components of Reynolds stresses (Fig.~\ref{flows}b), we can conclude that, even if the rotation effects are weak in the realistic rotation case, they still impact the dynamics of the convective flows below the leptocline.

\section{Conclusions}
The analysis of a 3D radiative hydrodynamic model of the 20-Mm upper layer of the solar convection zone confirms, by visual inspection of individual snapshots, previous findings of an increase with depth in the characteristic scales of convection down to about 7~Mm below the photosphere. On the other hand, the model reveals increasing complexity and homogeneity of the turbulent flows at greater depths. At those depths, analysis of the turbulent spectra is required to detect a continuing increase in convective scales with depth.

The qualitative change in the turbulent structure associated with the H and He-I ionization zone (which is co-located with the leptocline, a substructure of the Near-Subsurface Shear Layer in the upper convection zone) is most significantly manifested by variations in the density and temperature fluctuations. The demonstrated structural dependence of the turbulence on such flowfield properties is important for understanding their coupling throughout the convective interior, the solar surface, and the low atmosphere. More detailed studies are required, particularly using deeper computational domains and including magnetic fields.\\

{\bf Acknowledgements}\\
In this work, we used the Helioseismic and Magnetic Imager (HMI) global helioseismology data products onboard the NASA Solar Dynamics Observatory (SDO). The data have been obtained through the Joint Science Operations Center (JSOC). This work is supported by NASA Science DRIVE (Diversify, Realize, Integrate, Venture, Educate) Science Center ``Consequences of Fields and Flows in the Interior and Exterior of the Sun" (COFFIES) 80NSSC22M0162, NASA Heliophysics Supporting Research, and Heliophysics Guest Investigator Programs.
%\bibliography{conv_str}

\end{document}